\documentclass{article}
\usepackage[utf8]{inputenc}
\usepackage[a4paper, total={6.5in, 9in}]{geometry}
\usepackage{amsmath,amsfonts,amssymb}
\usepackage{graphicx}
\usepackage{hyperref}
\usepackage{subcaption}
\usepackage{mathrsfs}
 \usepackage{textcomp} 
\usepackage{pgfplots}
\usepackage{tikz}
\usetikzlibrary{
    arrows.meta,
    decorations.pathmorphing,
    positioning,
    calc,
    shapes.geometric
}
\usepackage{bm}
\newenvironment{acknowledgments}{\section*{Acknowledgments}}{}
\usepackage{float}
\usepackage{placeins}
\usepackage{multirow}
\usepackage{tikz}
\usetikzlibrary{trees,positioning}
\usepackage{forest}

\usepackage{booktabs}
\usepackage{siunitx}
\usepackage{xcolor}

\hypersetup{
  colorlinks=true,
  linkcolor=blue,
  citecolor=blue,
  urlcolor=blue
}

\newcommand{\vF}{v_{\mathrm F}}

\newcommand{\bk}{\bm{k}}
\newcommand{\bp}{\bm{p}}

\title{Kinematics of Inverse Compton Scattering from Dirac Carriers in Graphene}

\author{Dar Faheem Ul Haq, Naveed Ahmad Shah, SSZ Ashraf\\
Department of Physics\\
Faculty of Science\\
Aligarh Muslim University, Aligarh--202002, India\\
\texttt{ssz\_ashraf@rediffmail.com}}

\date{\today}

\begin{document}

\maketitle

\begin{abstract}
We examine the kinematics of Compton and inverse Compton
scattering by Dirac carriers in graphene. Unlike photons in
vacuum, which propagate at the speed of light $c$, graphene
carriers have a characteristic velocity $\vF$, with
$\vF\ll c$. This difference means that the standard Compton
scattering formula cannot be applied to graphene simply by
replacing $c$ with $\vF$. We first examine Compton-like
scattering for a gapless Dirac carrier initially at the Dirac
point and for a gapped carrier at the band edge, corresponding
to zero in-plane momentum. We then consider scattering from an
energetic gapless Dirac carrier,
\(\gamma+\mathrm{e}_{\rm D}
\longrightarrow
\gamma'+\mathrm{e}_{\rm D}'
\). By combining energy conservation with the graphene dispersion and in-plane momentum
conservation, we obtain a quadratic equation for the scattered
photon energy within the model considered. This relation is the
main kinematic result of the work. We identify inverse
Compton-like scattering by the transfer of energy from the
carrier to the photon, subject to the simultaneous conservation
of energy and in-plane momentum. We also discuss how this
process differs kinematically from Doppler-shifted emission,
Čerenkov-like radiation, harmonic generation, photon drag, and
other nonlinear optical processes. Finally, we propose a
time-resolved pump--probe arrangement in which the scattered
photon spectrum is the primary observable, while time- and
angle-resolved photoemission spectroscopy (TR-ARPES) provides
an independent probe of the nonequilibrium carrier
distribution. These results provide a kinematic framework for analysing inverse Compton-like scattering in graphene and for identifying the conditions under which energy can be transferred from energetic Dirac carriers to photons.
\end{abstract}

\section{Introduction}

Graphene offers a simple and intriguing setting in which to study
relativistic-like quasiparticles in condensed matter. Near its
inequivalent Dirac points, the low-energy electronic states of
monolayer graphene are described by an effective Dirac Hamiltonian.
In this description, the sublattice degree of freedom acts as a
pseudospin, and the energy varies linearly with the magnitude of
the wave vector:
\(
    E_{\bk}=\pm\hbar\vF|\bk|,
\)
where $\vF\simeq 10^6~\mathrm{m\,s^{-1}}$ is the characteristic
Dirac velocity
\cite{CastroNeto2009}.
This unusual electronic structure gives rise to several
interesting phenomena, including chiral transport, Klein
tunnelling, and Cerenkov-like emission
\cite{Katsnelson2006,Kaminer2016}.

The motion of carriers in graphene can also affect the frequency
and propagation of electromagnetic and collective excitations.
Doppler-related effects and magnetophonon phenomena have been
studied both theoretically and experimentally
\cite{Kim2014,Borgnia2015,Greenaway2021}. These effects are
related to carrier motion, but they should be distinguished from
photon--carrier scattering, in which an incident photon exchanges
energy and momentum with a carrier and emerges with a different
energy.

In conventional Compton scattering,
\(    \gamma+e^-\longrightarrow\gamma'+e^-,\)
an incident photon scatters from a free electron, transferring
energy to the electron in the standard case of an initially
stationary target. In inverse Compton scattering (ICS), an
energetic electron can instead transfer part of its energy to
the photon, producing a scattered photon of higher energy
\cite{Compton1923, Rindler2006, Arutyunian1963,Serafini2024}. The corresponding
energy-transfer conditions are
\(   E_{\gamma'}>E_{\gamma i}, E_f<E_i.
\)
These conditions describe the direction of energy transfer, but
they are not sufficient on their own to establish an
inverse-Compton process. The initial and final states must also
satisfy energy and momentum conservation.

The same problem cannot be treated in graphene simply by
replacing $c$ with $\vF$ in the vacuum Compton formula. Photons
obey the dispersion relation
\(    E_{\gamma i}=c|\bp_\gamma|,
\)
whereas a gapless Dirac carrier has the dispersion
\(E=\vF|\boldsymbol{\pi}|.\)
Since $\vF\ll c$, the two excitations have very different
characteristic velocities and obey different dispersion
relations. There is an additional constraint: carriers are
confined to the graphene plane, so their momentum balance
involves only the in-plane components of the photon momenta.
The scattering kinematics must therefore be derived directly
from the graphene dispersion and the relevant conservation laws.

In this work, we develop the kinematics of Compton-like and
inverse-Compton-like scattering by Dirac carriers in graphene.
We begin with two idealized limiting cases: a gapless carrier
at the Dirac point and a gapped carrier at the band edge,
corresponding to zero in-plane momentum. We then consider
scattering from an energetic gapless Dirac carrier,
\(\gamma+\mathrm{e}_{\rm D}
\longrightarrow
\gamma'+\mathrm{e}_{\rm D}',
\)
where $\mathrm{e}_{\rm D}$ and $\mathrm{e}_{\rm D}'$ denote the initial and final
Dirac carriers. By combining energy conservation with the
linear graphene dispersion and in-plane photon momentum
conservation, we obtain a quadratic equation for the scattered
photon energy within the model and geometry considered. This
equation is the principal kinematic result of the work.

An important part of the analysis is to distinguish the proposed
scattering channel from other processes that can produce
higher-frequency radiation. In inverse-Compton-like scattering,
the incident photon gains energy at the expense of the carrier,
and the initial and final states must satisfy both energy and
in-plane momentum conservation. Other mechanisms can produce
frequency shifts or up-conversion without involving this same
combination of photon scattering and carrier recoil \cite{Glazov2014,Hafez2018, Ornigotti2023}. Their main
characteristics are summarized in
Table~\ref{tab:upconversion_signatures}.

\begin{table}[htb!]
\centering
\renewcommand{\arraystretch}{1.2}
\begin{tabular}{|l|p{8.5cm}|}
\hline
\textbf{Process} & \textbf{Characteristic signature} \\
\hline
Inverse-Compton-like scattering &
An incident photon emerges with $\omega_f>\omega_i$ while the
carrier loses energy, $E_f<E_i$, subject to energy and
in-plane momentum conservation \cite{Arutyunian1963}.\\
\hline
Doppler-shifted emission &
The emitted frequency depends on carrier motion and observation
direction; an incident photon need not be involved\cite{Kim2014}.\\
\hline
Cerenkov-like radiation &
Emission is governed by phase matching and the relation between
the carrier velocity and the relevant mode phase velocity \cite{Kaminer2016}.\\
\hline
Harmonic generation &
Radiation is produced at harmonics of the driving frequency,
typically $\omega_f=n\omega_i$ for integer $n>1$; carrier recoil
is not its defining signature \cite{Glazov2014, Mikhailov2014}.\\
\hline
Other nonlinear up-conversion &
Higher-frequency radiation is generated through nonlinear
optical response, without necessarily involving the particular
photon--carrier recoil channel considered here\cite{Ornigotti2023}.\\
\hline
\end{tabular}
\caption{Kinematic distinctions between inverse-Compton-like
scattering and other mechanisms of frequency up-conversion.}
\label{tab:upconversion_signatures}
\end{table} 

The history of graphene research also shows why theoretical
predictions and experimental demonstrations need to be
distinguished. Several phenomena associated with graphene's
unusual electronic structure became experimentally accessible
as material quality, device fabrication, and measurement
techniques improved. Examples include the quantum Hall effect
and associated Berry phase \cite{Zhang2005}, Klein tunnelling
in graphene junctions \cite{Stander2009,YoungKim2009},
fractional quantum Hall states \cite{Bolotin2009,Du2009}, and
hydrodynamic electron transport
\cite{Crossno2016,Bandurin2016}. More recently, a mesoscopic
Klein--Schwinger effect was reported in ballistic graphene
transistors \cite{Schmitt2023}. These developments illustrate
the difference between identifying a theoretically possible
process and establishing its experimental manifestation.

Compton-profile measurements have already been performed on graphene,
but they probe electronic momentum distributions rather than the
photon--Dirac-carrier scattering process considered here. Feng
\textit{et al.}, for example, measured the valence Compton profile of
few-layer graphene using electron energy-loss spectroscopy and compared
the measured profile with full-potential electronic-structure
calculations \cite{Feng2019}. Such measurements demonstrate the
usefulness of Compton-profile techniques for probing graphene's
electronic structure, but they should be distinguished from inverse
Compton scattering, in which an incident photon gains energy from a
moving Dirac carrier.
Finally, we propose a time-resolved pump--probe geometry for
investigating the predicted scattering channel. The scattered
photon spectrum would be the primary observable, while
time- and angle-resolved photoemission spectroscopy (TR-ARPES)
could provide an independent measure of the nonequilibrium
carrier distribution\cite{Gierz2017,Ulstrup2015}. The present work establishes the
kinematic conditions for the proposed process; it does not
provide a fully normalized microscopic scattering rate or
demonstrate that an experimentally detectable up-converted
signal will result. Establishing the rate and assessing the
feasibility of its experimental observation remain important
steps for future work.

\section{Photon--Dirac-Carrier Kinematics}

\subsection{Graphene Dirac dispersion}

The Compton effect in graphene requires some care because the electronic
excitations in graphene are not ordinary massive relativistic electrons.
The low-energy electronic states near a Dirac point are described, within the ideal gapless Dirac approximation, by
\(
H_{\rm D}=\hbar \vF\,\boldsymbol{\sigma}\cdot\mathbf{k},
\)
whose eigenvalues are
\(
E_{\mathbf{k},s}=s\hbar \vF|\mathbf{k}|,
\ s=\pm1.
\)
Here, \(s=+1\) and \(s=-1\) denote the conduction and valence bands, respectively. In what follows, we consider the positive-energy branch,
\(
E_{\mathbf{k}}=\hbar \vF|\mathbf{k}|,
\)
where \(\vF\simeq 10^6\,{\rm m\,s^{-1}}\ll c\) is the Fermi velocity.~\cite{CastroNeto2009}
Thus, although the graphene quasiparticles exhibit a linear dispersion
resembling that of relativistic massless particles, their characteristic
velocity is \(\vF\), not the speed of light \(c\).

Consider the Compton-like process
\(\gamma+\mathrm{e}_{\rm D}
\longrightarrow
\gamma'+\mathrm{e}_{\rm D}',
\)
where \(\mathrm{e}_{\rm D}\) denotes a Dirac quasiparticle in graphene.
We denote the two-dimensional crystal momentum of a graphene
Dirac quasiparticle by $\boldsymbol{\pi}=\hbar\mathbf{k}$.
Its energy in the gapless Dirac approximation is
\(E=\vF|\boldsymbol{\pi}|.\)
It is useful to define a graphene pseudofour-momentum
\(
K^\mu
=
\left(
\frac{E}{\vF},
\boldsymbol{\pi}
\right).
\)
With the graphene Dirac metric
\(
\eta_{\mu\nu}
=
\operatorname{diag}(1,-1,-1),
\)
one obtains
\(
K^2
=
\frac{E^2}{\vF^2}
-\boldsymbol{\pi}^{\,2}
=0.
\)
Thus, in the effective Dirac description, a gapless graphene
quasiparticle is ``massless'' with respect to the effective
\(2+1\)-dimensional Dirac metric.
However, this effective Minkowski structure must not be confused with
physical spacetime. In particular,
\(\vF\neq c.\)
The photon four-momentum is null with respect to the electromagnetic
light cone determined by \(c\), whereas the graphene quasiparticle is
null with respect to the effective Dirac cone determined by \(\vF\).
Consequently, the usual single-metric four-momentum derivation of the
vacuum Compton formula does not apply directly to graphene.

\subsubsection{Energy and Momentum Conservation}

For a photon scattering from a graphene Dirac quasiparticle, energy
conservation requires
\begin{equation}
E_i+\hbar\omega_i=E_f+\hbar\omega_f.
\label{eq:photon_carrier_energy_conservation}
\end{equation}
In the translationally invariant graphene plane, momentum
conservation gives
\(
\boldsymbol{\pi}_i
+
\frac{\hbar \omega_i}{c}\mathbf{u}_i
=
\boldsymbol{\pi}_f
+
\frac{\hbar \omega_f}{c}\mathbf{u}_f.
\)
Here $\hat{\mathbf n}_i$ and $\hat{\mathbf n}_f$ are the unit vectors along the incident and scattered photon directions, and $\mathbf{u}_i$ and $\mathbf{u}_f$ are their respective in-plane projections. Only the component of the photon momentum parallel to the graphene
sheet participates in the in-plane momentum balance. The in-plane projection of the photon momentum is used solely for the momentum balance within the sheet; it should not be interpreted as a complete photon four-momentum in the effective graphene metric. If the photon
propagates within the graphene plane, then
\(\mathbf{u}_i=\hat{\mathbf n}_i,
\
\mathbf{u}_f=\hat{\mathbf n}_f.
\)
For a gapless Dirac quasiparticle,
\(
E_i=\vF|\boldsymbol{\pi}_i|,
\
E_f=\vF|\boldsymbol{\pi}_f|.
\)
Hence,
\(\vF|\boldsymbol{\pi}_i|+\hbar\omega_i
=\vF|\boldsymbol{\pi}_f|+\hbar\omega_f.\)
Using momentum conservation,
\(
\boldsymbol{\pi}_f
=
\boldsymbol{\pi}_i
+
\frac{\hbar \omega_i}{c}\hat{\mathbf n}_i
-
\frac{\hbar \omega_f}{c}\hat{\mathbf n}_f,
\)
we obtain
\begin{equation}  
\vF|\boldsymbol{\pi}_i|
+ \hbar \omega_i
=
\vF
\left|
\boldsymbol{\pi}_i
+
\frac{\hbar \omega_i}{c}\hat{\mathbf n}_i
-
\frac{\hbar \omega_f}{c}\hat{\mathbf n}_f
\right|
+\hbar \omega_f.
\end{equation}
This equation is the graphene analogue of the kinematic equation from
which the ordinary Compton shift is obtained.

\subsubsection{Special Case: Initial Dirac Point}

One might attempt to imitate the ordinary Compton derivation by taking
the initial graphene quasiparticle to have zero crystal momentum,
\(\mathbf{k}_i=0.\) The initial Dirac point is a zero-energy state in the ideal single-particle dispersion, not necessarily a freely available, occupied electron state that can undergo the assumed scattering event. The calculation establishes a kinematic solution, not the existence or measurable probability of a scattering process. Band occupancy, Pauli blocking, and the relevant interaction matrix elements would need to be considered for a physical prediction. For a gapless graphene spectrum, this corresponds to
\(E_i=0.\)

Momentum conservation then gives
\begin{equation}
\boldsymbol{\pi}_f=\frac{\hbar\omega_i}{c}\hat{\mathbf n}_i
-\frac{\hbar\omega_f}{c}\hat{\mathbf n}_f.
\label{eq:dirac_point_momentum}
\end{equation}
Consequently,
\(|\boldsymbol{\pi}_f|^2
=
\frac{\hbar^2}{c^2}
\left[
\omega_i^2+\omega_f^2
-2\omega_i \omega_f \cos\theta
\right].
\)
The final quasiparticle energy is therefore $E_f
= (\hbar \vF /c) (
\omega_i^2+\omega_f^2
-2\omega_i \omega_f \cos\theta
)^{1/2}$.

Energy conservation,
\(\hbar\omega_i=\hbar\omega_f+E_f,\)
then gives
\(
\hbar\omega_i
=
\hbar\omega_f
+
\frac{\hbar\vF}{c}
\sqrt{
\omega_i^2+\omega_f^2
-2\omega_i\omega_f\cos\theta
}.
\)
Equivalently,
\begin{equation}
(\omega_i-\omega_f)^2
=
\frac{\vF^2}{c^2}
\left(
\omega_i^2+\omega_f^2
-2\omega_i\omega_f\cos\theta
\right).
\end{equation}

Introducing
\(\beta_D=\frac{\vF}{c},\)
the resulting quadratic equation is
\begin{equation}
(1-\beta_D^2)\omega_f^2
-2\omega_i(1-\beta_D^2\cos\theta)\omega_f
+(1-\beta_D^2)\omega_i^2=0.
\label{eq:dirac_point_quadratic}
\end{equation}
Solving the quadratic equation yields two algebraic branches, corresponding to the positive and negative signs in the quadratic formula. The physically admissible branch must be identified from the kinematic conditions and the definitions of the scattering angle and Doppler parameter. For the branch considered here, we select the negative sign, obtaining
\begin{equation}
\omega_f
=
\omega_i\,
\frac{
1-\beta_D^2\cos\theta
-
\beta_D
\sqrt{
(1-\cos\theta)
\left[2-\beta_D^2(1+\cos\theta)\right]
}
}{
1-\beta_D^2
}.
\label{eq:dirac_point_relation}
\end{equation}

Equivalently, after rationalizing the expression,
\begin{equation}
E_{\gamma f}
=
\frac{
E_{\gamma i}(1-\beta_D^2)
}{
1-\beta_D^2\cos\theta
+
\beta_D
\sqrt{
(1-\cos\theta)
\left[2-\beta_D^2(1+\cos\theta)\right]
}
}.
\end{equation}

This expression may be compared with the ordinary Compton formula,
\(
E_{\gamma f}
=
\frac{E_{\gamma i}}
{
1+\dfrac{E_{\gamma i}}{mc^2}(1-\cos\theta)
}.
\)
The two expressions have fundamentally different origins. In ordinary
Compton scattering, the photon loses energy through recoil of a massive
particle with rest energy \(mc^2\). At the graphene Dirac point, in
contrast, the initial quasiparticle has zero energy and zero momentum,
and the result is governed by the velocity ratio
\(\beta_D=\frac{\vF}{c},\) 
rather than by a rest-mass scale.

Since
\(
\vF/c\sim1/300,
\)
the energy shift in this idealized Dirac-point process is small.
Moreover, because \(E_i=0\), the photon cannot gain energy from the
initial quasiparticle. Thus,
\(E_{\gamma f}<E_{\gamma i}\)
for nonzero scattering angle, and this process should be regarded as a
Compton-like energy-loss process rather than inverse Compton scattering.

\subsubsection{Why the Ordinary Compton Formula Does Not Follow}

The ordinary Compton formula is derived by combining relativistic
energy--momentum conservation with the mass-shell conditions for the
electron and the photons. For a free relativistic electron, the
four-momentum
\(P^\mu=\left(\frac{E}{c},\mathbf P\right)\)
satisfies
\(P^2=m^2c^2,\)
whereas the four-momenta \(p^\mu\) and \(p'^\mu\) of the incident
and scattered photons satisfy
\(p^2=p'^2=0.\)
Using these conditions together with four-momentum conservation,
\(P+p=P'+p',\)
we obtain the invariant relation
\(P\cdot(p-p')=p\cdot p',\)
which leads to the familiar Compton formula.

For graphene, however, the situation is somewhat different. Near a
Dirac point, the low-energy dispersion of a gapless graphene
quasiparticle is approximately linear:
\(E=\vF|\boldsymbol{\pi}|,\)
where \(\vF\ll c\) is the Fermi velocity and
\(\boldsymbol{\pi}\) is the in-plane crystal momentum measured
relative to the Dirac point. This dispersion resembles that of a
massless relativistic particle, but the characteristic velocity is
\(\vF\), not the speed of light.

To see why this distinction matters, suppose we construct a
four-component object using the physical velocity scale \(c\):
\(P^\mu=\left(\frac{E}{c},\boldsymbol{\pi}\right).\)
With the Minkowski metric of signature \((+,-,-,-)\), its squared
norm becomes
\begin{equation}
\begin{aligned}
P^2
=\frac{E^2}{c^2}-|\boldsymbol{\pi}|^2
=\left(\frac{\vF^2}{c^2}-1\right)
|\boldsymbol{\pi}|^2.
\end{aligned}
\end{equation}
Unlike the mass-shell condition for a free relativistic electron,
this expression depends on the quasiparticle momentum. It therefore
cannot be identified with a fixed rest-mass condition
\(P^2=m^2c^2\) for an ordinary relativistic particle. The
four-component object constructed above is consequently not the
physical four-momentum of a free electron transforming under the
vacuum Lorentz group.

There is, nevertheless, a useful way to express the graphene
dispersion in a relativistic-like form. Introducing the effective
pseudofour-momentum
\(K^\mu=\left(\frac{E}{\vF},\boldsymbol{\pi}\right),\)
we find
\begin{equation}
\begin{aligned}
K^2
=\frac{E^2}{\vF^2}-|\boldsymbol{\pi}|^2
=0,
\end{aligned}
\end{equation}
where the last equality follows directly from
\(E=\vF|\boldsymbol{\pi}|\). Thus, in the ideal low-energy Dirac
description, the gapless graphene quasiparticle lies on the null
cone defined by an effective metric with characteristic velocity
\(\vF\).

This result is mathematically similar to the null condition for a
vacuum photon, but the two should not be confused. The photon
dispersion in vacuum is governed by \(c\), whereas the graphene
quasiparticle dispersion is governed by \(\vF\). The two therefore
belong to different effective kinematic descriptions and cannot
simply be treated as massless excitations of a common Lorentzian
metric.

For this reason, the ordinary vacuum Compton derivation cannot be
applied directly to graphene by treating its Dirac quasiparticle as
a conventional relativistic electron. In particular, replacing
\(c\) with \(\vF\) in the standard Compton formula would also
incorrectly assign the graphene characteristic velocity to the
photon dispersion.

The appropriate approach is to formulate the scattering problem
using the actual dispersion relations and conservation laws for
both the photon and the graphene quasiparticle. For a process
involving one incident photon, one scattered photon, and a graphene
quasiparticle, energy conservation gives
\(E_i+\hbar\omega_i=E_f+\hbar\omega_f,\)
while conservation of in-plane momentum gives
\(\boldsymbol{\pi}_i+\hbar\mathbf{k}_{\gamma i,\parallel}
=
\boldsymbol{\pi}_f+\hbar\mathbf{k}_{\gamma f,\parallel}.
\)
Here, \(\mathbf{k}_{\gamma i,\parallel}\) and
\(\mathbf{k}_{\gamma f,\parallel}\) denote the in-plane components of the
incident and scattered photon wavevectors, respectively. These
equations must be solved together with the appropriate photon and
quasiparticle dispersion relations. If the lattice or another
scattering centre supplies or absorbs crystal momentum, the
corresponding momentum transfer must also be included.

The essential point is that photon scattering in graphene must be
analysed using the physical photon dispersion and the graphene
quasiparticle dispersion together. The resulting kinematics should
be derived from energy and momentum conservation, rather than by
directly importing the standard vacuum Compton invariant.

\subsection{Gapped Graphene}

The situation can be generalized to a gapped graphene system. If a
gap \(2\Delta\) is present, the quasiparticle dispersion becomes
\(E_{\boldsymbol{\pi}}
=
\sqrt{\Delta^2+\vF^2|\boldsymbol{\pi}|^2}.
\)
It is convenient to introduce the effective Dirac mass
\(m_D=\frac{\Delta}{\vF^2},\)
so that
\(E_{\boldsymbol{\pi}}^2=
m_D^2\vF^4+\vF^2|\boldsymbol{\pi}|^2.
\)
The corresponding effective pseudofour-momentum is
\(
K^\mu
=
\left(
\frac{E}{\vF},
\boldsymbol{\pi}
\right),
\)
which satisfies
\(K^2=m_D^2\vF^2,\)
where the effective metric is taken with signature
\((+,-,-)\) for the \(2+1\)-dimensional graphene quasiparticle
spacetime. Thus, a gapped graphene quasiparticle behaves as a massive
particle with respect to the effective Dirac metric. Nevertheless, the
characteristic velocity remains \(\vF\), whereas the electromagnetic
field propagates in vacuum with velocity \(c\). Consequently, the
ordinary vacuum Compton invariant cannot be transferred directly to
graphene, because the quasiparticle and electromagnetic field are
characterized by different invariant velocity scales.

The appropriate treatment must instead combine energy conservation,
\(E_i+\hbar\omega_i=E_f+\hbar\omega_f,\)
with conservation of the momentum component parallel to the graphene
plane,
\(\boldsymbol{\pi}_i+\mathbf{p}_{\gamma i,\parallel}
=
\boldsymbol{\pi}_f+\mathbf{p}_{\gamma f,\parallel},
\)
together with the graphene dispersion relation.

Consider a quasiparticle initially at rest in the graphene band,
\(\boldsymbol{\pi}_i=0,
\quad
E_i=\Delta.
\)
Let the incident and scattered photons have frequencies \(\omega_i\) and
\(\omega_f\), respectively. For photons propagating in the graphene plane,
their momenta are
\(\mathbf{p}_{\gamma i,\parallel}
=
\frac{\hbar\omega_i}{c}\,\hat{\mathbf n}_i,
\quad
\mathbf{p}_{\gamma f,\parallel}
=
\frac{\hbar\omega_f}{c}\,\hat{\mathbf n}_f,
\)
where \(\theta\) is the angle between
\(\hat{\mathbf n}_i\) and \(\hat{\mathbf n}_f\). Momentum conservation
therefore gives
\begin{equation}
\boldsymbol{\pi}_f
=
\mathbf{p}_{\gamma i,\parallel}
-
\mathbf{p}_{\gamma f,\parallel},
\end{equation}
and hence
\(|\boldsymbol{\pi}_f|^2
=
p_{\gamma i,\parallel}^2+p_{\gamma f,\parallel}^{\,2}
-2p_{\gamma i,\parallel}p_{\gamma f,\parallel}\cos\theta.
\)
Thus,
\(
|\boldsymbol{\pi}_f|^2
=
\left(\frac{\hbar\omega_i}{c}\right)^2
+
\left(\frac{\hbar\omega_f}{c}\right)^2
-
2\frac{\hbar^2\omega_i\omega_f}{c^2}\cos\theta.
\)

Using
\(E_f
=
\sqrt{\Delta^2+\vF^2|\boldsymbol{\pi}_f|^2}
\)
in the energy-conservation equation gives
\begin{equation}
\Delta+\hbar\omega_i
=
\sqrt{
\Delta^2+
\frac{\vF^2\hbar^2}{c^2}
\left[
\omega_i^2+\omega_f^2-2\omega_i\omega_f\cos\theta
\right]
}
+\hbar\omega_f.
\label{eq:gapped_energy_conservation}
\end{equation}
Rearranging and squaring Eq.~\eqref{eq:gapped_energy_conservation} yields
\(
2\Delta\hbar(\omega_i-\omega_f)
+
\hbar^2(\omega_i-\omega_f)^2
=
\frac{\vF^2\hbar^2}{c^2}
\left[
\omega_i^2+\omega_f^2-2\omega_i\omega_f\cos\theta
\right].
\)

Using \(\beta_D\), the resulting equation can be written as
\begin{equation}
(1-\beta_D^2)\omega_f^2
-
2\left[
\frac{\Delta}{\hbar}
+
\omega_i(1-\beta_D^2\cos\theta)
\right]\omega_f
+
\left[
2\frac{\Delta}{\hbar}\omega_i
+
(1-\beta_D^2)\omega_i^2
\right]
=0.
\label{eq:gapped_quadratic}
\end{equation}

The root continuously connected to the no-recoil limit is
\begin{equation}
\omega_f
=
\frac{
\displaystyle
\frac{\Delta}{\hbar}
+
\omega_i(1-\beta_D^2\cos\theta)
-
\sqrt{
\left[
\frac{\Delta}{\hbar}
+
\omega_i(1-\beta_D^2\cos\theta)
\right]^2
-
(1-\beta_D^2)
\left[
2\frac{\Delta}{\hbar}\omega_i
+
(1-\beta_D^2)\omega_i^2
\right]
}
}{
\displaystyle
1-\beta_D^2
}.
\end{equation}
The physical root is selected by requiring consistency with the original, unsquared energy-conservation equation and continuity with the no-recoil limit, \(\theta\to0\), for which \(\omega_f\to\omega_i\). The other algebraic root is inadmissible because it does not satisfy the original energy-conservation equation for the assumed initial state.

The result may also be expressed in terms of the photon energies
\(E_{\gamma i}=\hbar\omega_i,
\quad
E_{\gamma f}=\hbar\omega_f,
\)
as
\begin{equation}
E_{\gamma f}
=
\frac{
\displaystyle
\Delta
+
E_{\gamma i}(1-\beta_D^2\cos\theta)
-
\sqrt{
\left[
\Delta+E_{\gamma i}(1-\beta_D^2\cos\theta)
\right]^2
-
(1-\beta_D^2)
\left[
2\Delta E_{\gamma i}
+
(1-\beta_D^2)E_{\gamma i}^2
\right]
}
}{
\displaystyle
1-\beta_D^2
}.
\end{equation}
This expression is the graphene analogue of the Compton formula for a
quasiparticle initially at rest in a gapped Dirac band. Unlike the
ordinary vacuum Compton formula, it contains both \(\vF\) and \(c\),
reflecting the distinct kinematic structures of the graphene
quasiparticle and the electromagnetic field.

A useful consistency check is obtained by considering the formal limit
\(\vF\rightarrow c,
\quad
\beta_D^2\rightarrow1.
\)
Rather than substituting \(\beta_D^2=1\) directly into the closed-form
solution, one should take this limit in the quadratic relation above.
The quadratic term then vanishes, giving
\(
2\Delta\hbar(\omega_i-\omega_f)
+
\hbar^2(\omega_i-\omega_f)^2
=
\hbar^2
\left[
\omega_i^2+\omega_f^2-2\omega_i\omega_f\cos\theta
\right].
\)
After simplification,
\(2\Delta(\omega_i-\omega_f)
=
2\hbar\omega_i\omega_f(1-\cos\theta),
\)
Identifying
\(\Delta=mc^2,
\)
one obtains
\(\lambda'-\lambda
=
\frac{2\pi\hbar}{mc}(1-\cos\theta),
\)
which is the standard Compton wavelength shift.

\subsubsection{Physical Interpretation}

For gapless and gapped graphene in the low-energy Dirac regime, the
distinction between an ordinary relativistic electron and graphene
Dirac quasiparticles can be summarized as follows:

\begin{table}[htbp]
\centering
\renewcommand{\arraystretch}{1.35}
\begin{tabular}{|l|c|c|c|}
\hline
\textbf{Property}
& \textbf{Electron}
& \textbf{Gapless graphene}
& \textbf{Gapped graphene}
\\
\hline

Dispersion
&
\(E^2=m^2c^4+c^2|\mathbf P|^2\)
&
\(E^2=v_F^2|\boldsymbol{\pi}|^2\)
&
\(E^2=\Delta^2+v_F^2|\boldsymbol{\pi}|^2\)
\\
\hline

Characteristic velocity
&
\(c\)
&
\(v_F\)
&
\(v_F\)
\\
\hline

Effective mass
&
\(m\)
&
\(0\)
&
\(m_D=\dfrac{\Delta}{v_F^2}\)
\\
\hline

Four-momentum
&
\(P^\mu=\left(\dfrac{E}{c},\mathbf P\right)\)
&
\(K^\mu=\left(\dfrac{E}{v_F},\boldsymbol{\pi}\right)\)
&
\(K^\mu=\left(\dfrac{E}{v_F},\boldsymbol{\pi}\right)\)
\\
\hline

Mass shell
&
\(P^\mu P_\mu=m^2c^2\)
&
\(K^\mu K_\mu=0\)
&
\(K^\mu K_\mu=m_D^2v_F^2\)
\\
\hline

Momentum conservation
&
\(\mathbf P+\mathbf p_\gamma
=\mathbf P'+\mathbf p_\gamma'\)
&
\(\boldsymbol{\pi}+\mathbf p_{\gamma\parallel}
=\boldsymbol{\pi}'+\mathbf p'_{\gamma\parallel}\)
&
\(\boldsymbol{\pi}+\mathbf p_{\gamma\parallel}
=\boldsymbol{\pi}'+\mathbf p'_{\gamma\parallel}\)
\\
\hline

Effective symmetry
&
\(c\)-Lorentz
&
\(v_F\)-Lorentz-like
&
\(v_F\)-Lorentz-like
\\
\hline

\end{tabular}

\caption{Comparison of the kinematic properties of an ordinary
relativistic electron with gapless and gapped graphene Dirac
quasiparticles. The graphene quantities refer to the low-energy Dirac
description, with \(v_F\) as the characteristic Dirac velocity and
\(m_D=\Delta/v_F^2\) as the effective Dirac mass in the gapped case.
For graphene, momentum conservation applies to the in-plane
components of the photon momenta,
\(\mathbf p_{\gamma,\parallel}\) and
\(\mathbf p'_{\gamma,\parallel}\).}
\label{tab:electron_graphene_comparison}
\end{table}

The table highlights the distinction between the effective Dirac
kinematics of graphene quasiparticles and the physical relativistic
kinematics of vacuum electrons and photons. In gapped graphene, the
dispersion resembles that of a massive relativistic particle, with
effective mass \(m_D=\Delta/\vF^2\); the gapless case corresponds to
the massless limit of the same effective description. However, the
characteristic velocity of the graphene quasiparticle is \(\vF\),
whereas the photon propagates in vacuum at \(c\). Since \(\vF\ll c\),
the two excitations do not share a common Lorentz kinematic structure.
Consequently, the ordinary vacuum Compton formula cannot be obtained
simply by replacing \(c\) with \(\vF\). The scattered-photon energy
must instead be determined from the respective dispersion relations,
energy conservation, and conservation of in-plane momentum.

For an initially stationary, gapless Dirac quasiparticle, the
simultaneous conservation of energy and in-plane momentum admits a
nontrivial lower solution for the scattered-photon energy at nonzero
scattering angle. In the idealized kinematic model considered above,
this solution satisfies \(E_{\gamma f}<E_{\gamma i}\), indicating energy
transfer from the photon to the quasiparticle. This result describes
kinematic compatibility rather than establishing an experimentally
observable scattering rate.

The observability of such a process depends additionally on electronic
state occupations, Pauli blocking, and the relevant interaction
matrix elements. Depending on the experimental geometry and material
conditions, effects such as interband transitions, phonons, disorder,
and substrate coupling may also need to be included in a microscopic
description of photon scattering in graphene.

\section{Inverse Compton Scattering in Graphene}
\label{sec:inverse_compton_graphene}

The possibility of realizing a Compton-like photon--carrier scattering
process in graphene is interesting because graphene supports Dirac
quasiparticles with a linear energy--momentum dispersion. However, the
kinematics of these quasiparticles differ fundamentally from those of
free relativistic electrons in vacuum. In particular, the characteristic
velocity of a graphene Dirac quasiparticle is the Fermi velocity
\(\vF\simeq 10^6\,{\rm m\,s^{-1}}\), whereas electromagnetic photons
propagate at the speed of light \(c\).

This section develops the kinematic relations for photon scattering
from gapless graphene Dirac quasiparticles. It distinguishes
Compton-like scattering from inverse Compton scattering and derives
the scattered-photon energy using energy conservation, in-plane
momentum conservation, and the linear graphene dispersion.

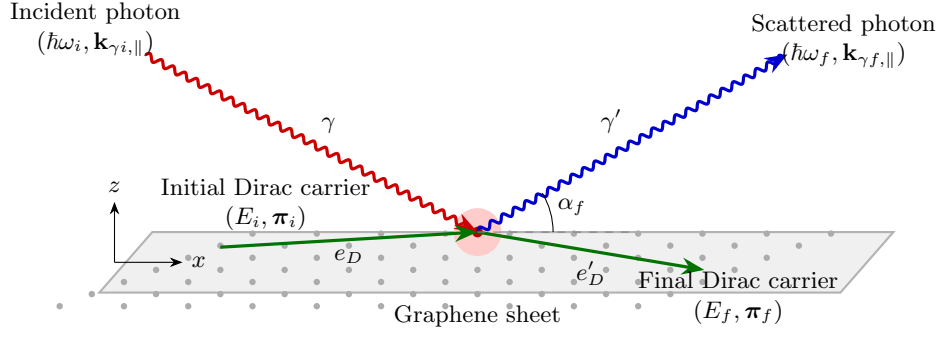
\begin{figure}[h!]
\centering
\begin{tikzpicture}[
    >=Stealth,
    photon/.style={
        -{Stealth[length=3mm]},
        very thick,
        decorate,
        decoration={
            snake,
            amplitude=1.2pt,
            segment length=5pt
        }
    },
    carrier/.style={
        -{Stealth[length=3mm]},
        very thick
    },
    every node/.style={font=\small}
]

\fill[gray!12]
    (-5,-0.65) -- (4.8,-0.65) --
    (5.5,0.15) -- (-4.3,0.15) -- cycle;

\draw[gray!60,thick]
    (-5,-0.65) -- (4.8,-0.65) --
    (5.5,0.15) -- (-4.3,0.15) -- cycle;

\foreach \j in {-3,...,3}{
    \foreach \i in {-5,...,4}{
        \coordinate (A) at
        ({0.85*\i+0.425*\j},{-0.35+0.16*\j});
        \fill[gray!65] (A) circle (1.2pt);
    }
}

\node[below] at (0,-0.7) {Graphene sheet};

\coordinate (O) at (0,0.15);
\fill[red!20] (O) circle (9pt);
\fill[red!70!black] (O) circle (2pt);

\draw[photon,red!80!black] (-4.4,2.5) -- (O)
    node[midway,above right,black] {$\gamma$};

\node[above left,align=center] at (-3.8,2.35)
    {Incident photon\\
    $(\hbar\omega_i,\mathbf{k}_{\gamma i,\parallel})$};

\draw[photon,blue!80!black] (O) -- (4.1,2.5)
    node[midway,above left,black] {$\gamma'$};

\node[above right,align=center] at (3.5,2.2)
    {Scattered photon\\
    $(\hbar\omega_f,\mathbf{k}_{\gamma f,\parallel})$};

\draw[carrier,green!45!black] (-3.4,-0.05) -- (O)
    node[midway,below,black] {$e_D$};

\node[above,align=center] at (-2.8,0.05)
    {Initial Dirac carrier\\
    $(E_i,\boldsymbol{\pi}_i)$};

\draw[carrier,green!45!black] (O) -- (3.0,-0.35)
    node[midway,below,black] {$e'_D$};

\node[below right,align=center] at (2.0,-0.25)
    {Final Dirac carrier\\
    $(E_f,\boldsymbol{\pi}_f)$};

\draw[dashed,gray] (O) -- (2.0,0.15);

\draw (1.0,0.15)
    arc[start angle=0,end angle=31,radius=1.0];
\node at (1.25,0.48) {$\alpha_f$};

\draw[->,thin] (-4.8,-0.25) -- (-3.9,-0.25)
    node[right] {$x$};
\draw[->,thin] (-4.8,-0.25) -- (-4.8,0.55)
    node[above] {$z$};

\end{tikzpicture}

\caption{Schematic ray diagram of photon scattering by Dirac carriers
in graphene. For the proposed inverse-Compton channel,
$\omega_f>\omega_i$ and $E_f<E_i$, subject to energy and in-plane
momentum conservation. The angle $\alpha_f$ shown in the diagram is measured from the indicated in-plane reference direction. The diagram is schematic and not to scale.}
\label{fig:graphene_dirac_ics_rays}
\end{figure}

\subsection{Dirac quasiparticles and energy conservation}

Near a Dirac point, the low-energy electronic excitations of monolayer
graphene are described by the effective Hamiltonian
\(H_{\rm D}
=
\hbar \vF\boldsymbol{\sigma}\cdot\mathbf{k},
\)
where \(\boldsymbol{\sigma}\) denotes the Pauli matrices acting on
the sublattice pseudospin degree of freedom. The corresponding energy
dispersion is
\(E_{\mathbf{k}}=\hbar \vF|\mathbf{k}|.
\) In the following kinematic treatment, \(E_{\mathbf k}=\hbar \vF|\mathbf k|\) denotes the positive-energy branch, and the initial and final states are taken to be conduction-band quasiparticle states. The occupation of the relevant electronic states must be considered separately when assessing whether a transition is physically allowed.

Consider a graphene Dirac quasiparticle with initial wave vector
\(\mathbf{k}_i\), energy
\(E_i=\hbar \vF k_i,\)
and an incident photon of frequency \(\omega_i\) and energy
\(E_{\gamma i}=\hbar\omega_i.\)
After scattering, the photon has frequency \(\omega_f\) and energy
\(E_{\gamma f}=\hbar\omega_f,\)
while the quasiparticle has final wave vector \(\mathbf{k}_f\) and
energy
\(E_f=\hbar \vF k_f.\)

The scattering process is represented schematically by
\(\gamma+\mathrm{e}_{\rm D}
\longrightarrow
\gamma'+\mathrm{e}_{\rm D}',
\)
where $\mathrm{e}_{\rm D}$ and $\mathrm{e}_{\rm D}'$ denote the
initial and final graphene Dirac quasiparticle states, respectively,
as illustrated in Fig.~\ref{fig:graphene_dirac_ics_rays}.

Energy conservation requires
\(E_i+E_{\gamma i}=E_f+E_{\gamma f},\)
or equivalently,
\(E_f=E_i+E_{\gamma i}-E_{\gamma f}.\)
Consequently,
\(\hbar(\omega_f-\omega_i)=E_i-E_f.\)

Thus, photon up-conversion, defined by
\(\omega_f>\omega_i,\)
requires
\(E_f<E_i.\)
The photon gains energy at the expense of the electronic carrier.
This energy-transfer condition is necessary for an inverse-Compton
interpretation, but it is not sufficient by itself to establish
inverse Compton scattering.

\subsection{In-plane momentum conservation}

The incident and scattered photons propagate in three-dimensional
space, with momenta
\(
\mathbf{p}_{\gamma i}
=
\frac{\hbar\omega_i}{c}\hat{\mathbf n}_i,
\quad
\mathbf{p}_{\gamma f}
=
\frac{\hbar\omega_f}{c}\hat{\mathbf n}_f.
\)
Here \(\hat{\mathbf n}_i\) and \(\hat{\mathbf n}_f\) are unit vectors
in the incident and scattered photon directions, respectively.

Since the electronic degrees of freedom are confined to the graphene
plane, the relevant photon momenta in the electronic momentum balance
are their in-plane projections. Let \(\mathsf P_\parallel\) denote
the projection operator onto the graphene plane and define
\(
\mathbf{u}_i=\mathsf P_\parallel\hat{\mathbf n}_i,
\quad
\mathbf{u}_f=\mathsf P_\parallel\hat{\mathbf n}_f.
\)
In general,
\(|\mathbf{u}_i|\leq 1,
\quad
|\mathbf{u}_f|\leq 1.
\)
These projected vectors are not necessarily unit vectors.

For a normal process, neglecting Umklapp scattering, conservation
of in-plane crystal momentum gives
\begin{equation}
\hbar\mathbf{k}_i+
\frac{\hbar\omega_i}{c}\mathbf{u}_i
=
\hbar\mathbf{k}_f+
\frac{\hbar\omega_f}{c}\mathbf{u}_f.    
\end{equation}
Hence,
\(\mathbf{k}_f
=
\mathbf{k}_i+
\frac{\omega_i}{c}\mathbf{u}_i
-
\frac{\omega_f}{c}\mathbf{u}_f.
\)

More generally, crystal momentum is conserved modulo a reciprocal
lattice vector \(\mathbf G\):
\(
\hbar\mathbf{k}_i+
\frac{\hbar\omega_i}{c}\mathbf{u}_i
=
\hbar\mathbf{k}_f+
\frac{\hbar\omega_f}{c}\mathbf{u}_f+
\hbar\mathbf{G}.
\)
The subsequent derivation concerns the normal process, for which
\(\mathbf G=0\).

For compactness, define the dimensionless geometrical factors
\(
s_i=|\mathbf{u}_i|^2,
\quad
s_f=|\mathbf{u}_f|^2,
\)
\(
d_i=\hat{\mathbf k}_i\cdot\mathbf{u}_i,
\quad
d_f=\hat{\mathbf k}_i\cdot\mathbf{u}_f,
\quad
d_{if}=\mathbf{u}_i\cdot\mathbf{u}_f,
\)
where \(\hat{\mathbf k}_i=\mathbf{k}_i/k_i\) for \(k_i\neq0\).
These factors account explicitly for the photon propagation geometry.
When both photon momenta lie in the graphene plane,
\(s_i=s_f=1\), and the geometrical factors can be expressed in terms
of ordinary in-plane scattering angles.

\subsection{Derivation of the scattered-photon energy}

Using the linear dispersion and the conservation law in
Eq.~\eqref{eq:photon_carrier_energy_conservation}, we obtain
\begin{equation}
k_f=k_i+\frac{\omega_i-\omega_f}{\vF}.
\label{eq:ics_kf_energy}
\end{equation}

Squaring the in-plane momentum-conservation equation yields
\begin{equation}
\begin{aligned}
k_f^2={}&k_i^2+\frac{\omega_i^2}{c^2}s_i
+\frac{\omega_f^2}{c^2}s_f
+\frac{2k_i\omega_i}{c}d_i-\frac{2k_i\omega_f}{c}d_f
-\frac{2\omega_i\omega_f}{c^2}d_{if}.
\end{aligned}
\label{eq:ics_kf_momentum}
\end{equation}

Substituting Eq.~\eqref{eq:ics_kf_energy} into
Eq.~\eqref{eq:ics_kf_momentum}, and using
\(
E_i=\hbar \vF k_i,\quad
E_{\gamma i}=\hbar\omega_i,\quad
E_{\gamma f}=\hbar\omega_f,
\)
gives the quadratic equation
\begin{equation}
(1-\beta_D^2s_f)E_{\gamma f}^2
-2\left[
E_i(1-\beta_Dd_f)
+E_{\gamma i}(1-\beta_D^2d_{if})
\right]E_{\gamma f}
+E_{\gamma i}\left[
2E_i(1-\beta_D d_i)
+E_{\gamma i}(1-\beta_D^2s_i)
\right]=0.
\label{eq:graphene_ics_quadratic}
\end{equation}

Define
\(A=
E_i(1-\beta_D d_f)+E_{\gamma i}(1-\beta_D^2d_{if})
\)
and
\(B=
E_{\gamma i}\left[
2E_i(1-\beta_D d_i)+E_{\gamma i}(1-\beta_D^2s_i)
\right].
\)
The two algebraic roots of Eq.~\eqref{eq:graphene_ics_quadratic} are
\begin{equation}
 E_{\gamma f,\pm}
=
\frac{
A\pm\sqrt{A^2-(1-\beta_D^2s_f)B}
}{
1-\beta_D^2s_f
}.   
\end{equation}
For the low-photon-energy branch, the physical solution is
\begin{equation}
E_{\gamma f}
=
\frac{
A-\sqrt{A^2-(1-\beta_D^2s_f)B}
}{
1-\beta_D^2s_f
}.
\label{eq:graphene_ics_solution}
\end{equation}
For \(E_{\gamma i}/E_i\ll1\), this branch satisfies
\begin{equation}
\frac{E_{\gamma f}}{E_{\gamma i}}
\longrightarrow
\frac{1-\beta_D d_i}{1-\beta_D d_f}.    
\end{equation}
The selected solution must also satisfy \(E_{\gamma f}>0\) and the
original, unsquared energy- and momentum-conservation equations.
The minus sign is not a universal root-selection rule for arbitrary
kinematic problems; its physical admissibility must be checked for
the scattering regime under consideration.

For inverse Compton scattering, the physically admissible solution must satisfy \(E_{\gamma f}>E_{\gamma i}\), or equivalently \(E_f<E_i\). Both algebraic roots should therefore be examined when identifying the relevant scattering branch. Each candidate must have a non-negative discriminant, a positive scattered-photon energy, and a final carrier energy \(E_f=E_i+E_{\gamma i}-E_{\gamma f}\geq0\), and must satisfy the original energy- and momentum-conservation equations. The lower root describes the branch for which \(E_{\gamma f}/E_{\gamma i}\) remains finite in the limit \(E_{\gamma i}/E_i\to0\); it should not be assumed to represent every physically admissible inverse-Compton process.

\subsection{Dimensionless form}

To make the dependence on the incident carrier and photon energies
explicit, define
\(\epsilon=\frac{E_{\gamma i}}{E_i},
\quad
x=\frac{E_{\gamma f}}{E_{\gamma i}}.
\)
For \(E_i>0\), the quadratic equation becomes
\begin{equation}
\epsilon(1-\beta_D^2s_f)x^2-2\left[
1-\beta_D d_f+\epsilon(1-\beta_D^2d_{if})
\right]x+2(1-\beta_D d_i)+\epsilon(1-\beta_D^2s_i)=0.
\end{equation}
Define
\(
A_D=1-\beta_D d_f+\epsilon(1-\beta_D^2d_{if})
\)
and
\(
C_D=2(1-\beta_D d_i)+\epsilon(1-\beta_D^2s_i).
\)
The low-photon-energy branch of Eq.~\eqref{eq:graphene_ics_quadratic} is then
\begin{equation}
 x=
\frac{
A_D-\sqrt{
A_D^2-\epsilon(1-\beta_D^2s_f)C_D
}
}{
\epsilon(1-\beta_D^2s_f)
}.   
\end{equation}
The scattered-photon energy follows from
\(E_{\gamma f}=xE_{\gamma i}.\)
This dimensionless representation is useful for exploring the
dependence of the energy transfer on the carrier energy, the
photon energy, the Fermi velocity, and the scattering geometry.

\subsection{The Dirac-point limit}

Consider the special case of an initial quasiparticle at the Dirac
point, \(E_i=0\). A particularly simple expression is obtained when
both photon momenta lie in the graphene plane. In that geometry,
\(s_i=s_f=1,\quad
d_{if}=\cos\theta,\)
where \(\theta\) is the angle between the incident and scattered
photon directions in the plane.

The quadratic equation reduces to
\begin{equation}
(1-\beta_D^2)E_{\gamma f}^2
-2E_{\gamma i}(1-\beta_D^2\cos\theta)E_{\gamma f}
+(1-\beta_D^2)E_{\gamma i}^2=0.    
\end{equation}
Its lower root is
\begin{equation}
E_{\gamma f}
=
E_{\gamma i}
\frac{
1-\beta_D^2\cos\theta
-\beta_D\sqrt{
(1-\cos\theta)
[2-\beta_D^2(1+\cos\theta)]
}
}{
1-\beta_D^2
}.
\end{equation}
Rationalizing the numerator gives the equivalent expression
\begin{equation}
E_{\gamma f}
=
\frac{
E_{\gamma i}(1-\beta_D^2)
}{
1-\beta_D^2\cos\theta
+\beta_D\sqrt{
(1-\cos\theta)
[2-\beta_D^2(1+\cos\theta)]
}
}.
\end{equation}

For \(0<\beta_D<1\) and nonzero scattering angle, this branch satisfies
\(E_{\gamma f}<E_{\gamma i}.\)
The photon therefore loses energy in this special kinematic limit.
It must not be identified with inverse Compton scattering, which
requires energy transfer from an energetic carrier to the photon.
Moreover, the Dirac-point limit alone does not establish a measurable
scattering process; the electronic occupation, available final
states, and interaction matrix elements must also be considered.

\subsection{Effective interaction}

The coupling of graphene Dirac quasiparticles to an electromagnetic
field can be introduced through minimal coupling. Taking \(e>0\)
to denote the magnitude of the electron charge, one uses
\(\boldsymbol{\pi}\longrightarrow
\boldsymbol{\pi}+e\mathbf A_\parallel,
\)
where \(\mathbf A_\parallel\) is the vector potential projected
onto the graphene plane. The effective Hamiltonian becomes
\(
H=\vF\boldsymbol{\sigma}\cdot
(\boldsymbol{\pi}+e\mathbf A_\parallel).
\)
Separating the unperturbed and interaction terms gives
\(H=H_0+H_{\rm int},
\)
with
\(H_0=\vF\boldsymbol{\sigma}\cdot\boldsymbol{\pi},
\quad
H_{\rm int}=e\vF
\boldsymbol{\sigma}\cdot\mathbf A_\parallel.
\)
The photon--carrier scattering amplitude consequently depends on
the Dirac pseudospin matrix elements, the photon polarization, and
the electronic states involved. Kinematic compatibility is
necessary, but it does not determine the scattering rate or
establish experimental observability.

\subsection{Comparison with ordinary Compton scattering}

For a photon scattering from an initially stationary free electron,
the ordinary Compton relation is
\(E_{\gamma f}
= E_{\gamma i} \left(
1+\dfrac{E_{\gamma i}}{m_ec^2}(1-\cos\theta) \right)^{-1}.
\)
This result follows from energy--momentum conservation together with
the relativistic electron dispersion
\(E_e^2=m_e^2c^4+c^2p_e^2.\)

For an energetic free electron, the inverse-Compton kinematics can
be written as
\begin{equation}
 E_{\gamma f}
=
\frac{
E_{\gamma i}(1-\beta\cos\theta_i)
}{
1-\beta\cos\theta_f+
\dfrac{E_{\gamma i}}{E_e}(1-\cos\theta)
},   
\end{equation}
where \(\beta=v/c\), \(E_e\) is the initial electron energy,
\(\theta_i\) and \(\theta_f\) are the angles between the electron
momentum and the incident and scattered photon directions, and
\(\theta\) is the angle between the two photon directions.

The graphene relations derived above are different because the
electronic dispersion is
\(E=\vF|\boldsymbol{\pi}|,\)
whereas the photon dispersion is
\(E_{\gamma i}=cp_\gamma.\)
The carrier and photon therefore involve different characteristic
velocities. In particular, replacing \(c\) by \(\vF\) in a vacuum
Compton formula does not yield the graphene scattering relation.

For photons propagating in the graphene plane, the graphene
kinematic equation is quadratic in \(E_{\gamma f}\), with coefficient
\(1-\beta_D^2\). Its formal linearization as \(\beta_D\to 1\) is an algebraic
limit and does not, by itself, establish recovery of the ordinary
massive-electron Compton formula. A physical reduction to that
formula would additionally require the appropriate dispersion
relation and identification of the rest-energy scale.

\begin{table}[h!]
\centering
\renewcommand{\arraystretch}{1.5}
\small
\begin{tabular}{|p{2.2cm}|p{11.3cm}|}
\hline
\textbf{Process} & \textbf{Photon-energy relation}\\
\hline
\textbf{Ordinary Compton scattering}
&
For a photon scattering from an initially stationary electron,
\[
\displaystyle
E_\gamma'=
\frac{E_\gamma}{
1+\dfrac{E_\gamma}{m_ec^2}(1-\cos\theta)}
\]
\\
\hline
\textbf{Compton-like scattering in gapless graphene}
&
For an initial Dirac-point quasiparticle, \(E_i=0\), with both
photon momenta in the graphene plane:
\[
\displaystyle
E_\gamma'=
\frac{
E_\gamma(1-\beta_D^2)
}{
1-\beta_D^2\cos\theta
+
\beta_D
\sqrt{
(1-\cos\theta)
[2-\beta_D^2(1+\cos\theta)]
}
}.
\]
\\
\hline
\textbf{Compton-like scattering in gapped graphene}
&
For a quasiparticle initially at the band edge, \(E_i=\Delta\),
with both photon momenta in the graphene plane:
\[
\displaystyle
E_\gamma'=
\frac{
A_g-
\sqrt{
A_g^2-(1-\beta_D^2)E_\gamma
\left[
2\Delta+E_\gamma(1-\beta_D^2)
\right]
}
}{
1-\beta_D^2
},
\]
where \quad
\(
\displaystyle
A_g=\Delta+E_\gamma(1-\beta_D^2\cos\theta).
\)
\\
\hline
\textbf{Photon scattering by an energetic gapless graphene quasiparticle}
&
For an energetic Dirac quasiparticle, \(E_i=\hbar v_F k_i\).
Then
\[
\displaystyle
E_\gamma'=
\frac{
A-
\sqrt{
A^2-(1-\beta_D^2s_f)B
}
}{
1-\beta_D^2s_f
},
\]
where
\[
\begin{aligned}
A={}&E_i(1-\beta_D d_f)+E_\gamma(1-\beta_D^2d_{if}),
\\[2pt]
B={}&E_\gamma\left[
2E_i(1-\beta_D d_i)
+E_\gamma(1-\beta_D^2s_i)
\right].
\end{aligned}
\]
The geometrical factors \(s_i,s_f,d_i,d_f,d_{if}\) are defined
in the text. The expression gives the lower algebraic root of the kinematic equation. It represents inverse-Compton-type energy transfer only when the selected root satisfies \(E_\gamma'>E_\gamma\), together with the original conservation equations and the physical admissibility conditions. Otherwise, it describes a different kinematic branch of photon scattering.
\\
\hline
\textbf{Ordinary inverse Compton scattering}
&
For an ordinary relativistic electron,
\[
\displaystyle
E_\gamma'=
\frac{
E_\gamma(1-\beta\cos\theta_i)
}{
1-\beta\cos\theta_f
+\dfrac{E_\gamma}{E_e}(1-\cos\theta)
},
\]
where \(E_e\) is the initial total electron energy,
and \(\theta\) is the angle between the incident and scattered
photon directions.
\\
\hline
\end{tabular}
\caption{Comparison of photon-energy relations for ordinary Compton
scattering, Compton-like scattering in gapless and gapped graphene,
and inverse Compton scattering from energetic carriers. For graphene,
\(\beta_D=v_F/c\); for an ordinary relativistic electron,
\(\beta=v/c\). The gapless Dirac-point and gapped-graphene expressions
assume coplanar photon propagation and normal in-plane momentum
conservation. The gapped expression assumes an initially stationary
quasiparticle at the band edge, \(E_i=\Delta\). The table summarizes
kinematic relations rather than scattering rates or evidence of
experimental observation.}
\label{tab:compton_graphene_comparison}
\end{table}

\paragraph{Scope and limitations.}
The present treatment focuses on the kinematics of photon scattering
from gapless graphene Dirac quasiparticles. A quantitative prediction
for inverse Compton scattering requires the initial electronic
distribution, photon polarization and propagation geometry,
pseudospin-dependent interaction matrix elements, and the associated
scattering probabilities. The kinematic equations identify the
energy and momentum combinations permitted by the stated assumptions,
but do not establish the magnitude or experimental observability of
the process. Gapped graphene has a different dispersion,
\(E=\sqrt{\Delta^2+\hbar^2\vF^2k^2},
\) and its scattering kinematics must account for the gap energy scale,
as derived in the preceding subsection.

\section{Proposed Experimental Geometry for Inverse Compton Scattering in Graphene}
\label{sec:experimental_geometry_ics}

An experimental study of inverse Compton scattering (ICS) in graphene
would require a graphene device, a population of energetic carriers,
and a controlled incident photon beam. The carriers could be accelerated
by an in-plane electric field or generated using an ultrafast optical
pump. Optical excitation can produce a nonequilibrium carrier
distribution that evolves through carrier--carrier and carrier--phonon
scattering \cite{Malic2017}. A probe photon beam of frequency
$\omega_i$ could then be directed onto the graphene sheet, with its
incidence angle and polarization varied to investigate their effects
on the scattering process.

The scattered photons would be detected with sufficient spectral and
angular resolution to determine their energies $\hbar\omega_f$ and
wave vectors $\mathbf{k}_{\gamma f}$. This measurement would provide the main
evidence for the proposed ICS process. Where feasible, time- and
angle-resolved photoemission spectroscopy (TR-ARPES) could be used
alongside photon detection to examine the carrier distribution before
and after the interaction. This technique has already been used to
study the evolution of photoexcited carriers in graphene
\cite{Gierz2017}.

In TR-ARPES, a probe pulse ejects some of the carriers from the
graphene sheet. Their kinetic energies and emission angles are then
measured to reconstruct the electronic energy--momentum distribution.
These photoelectrons serve as a diagnostic signal; they are not
products of the ICS process. TR-ARPES therefore provides information
about the carrier population as a whole rather than tracking the
recoil of an individual carrier. Pump-induced space-charge and
surface-photovoltage effects must also be taken into account when
interpreting the measurements \cite{Ulstrup2015}.

\begin{figure}[h!]
    \centering
    \IfFileExists{ics_trarpes_schematic_pi_updated.pdf}{%
    \includegraphics[width=\textwidth]{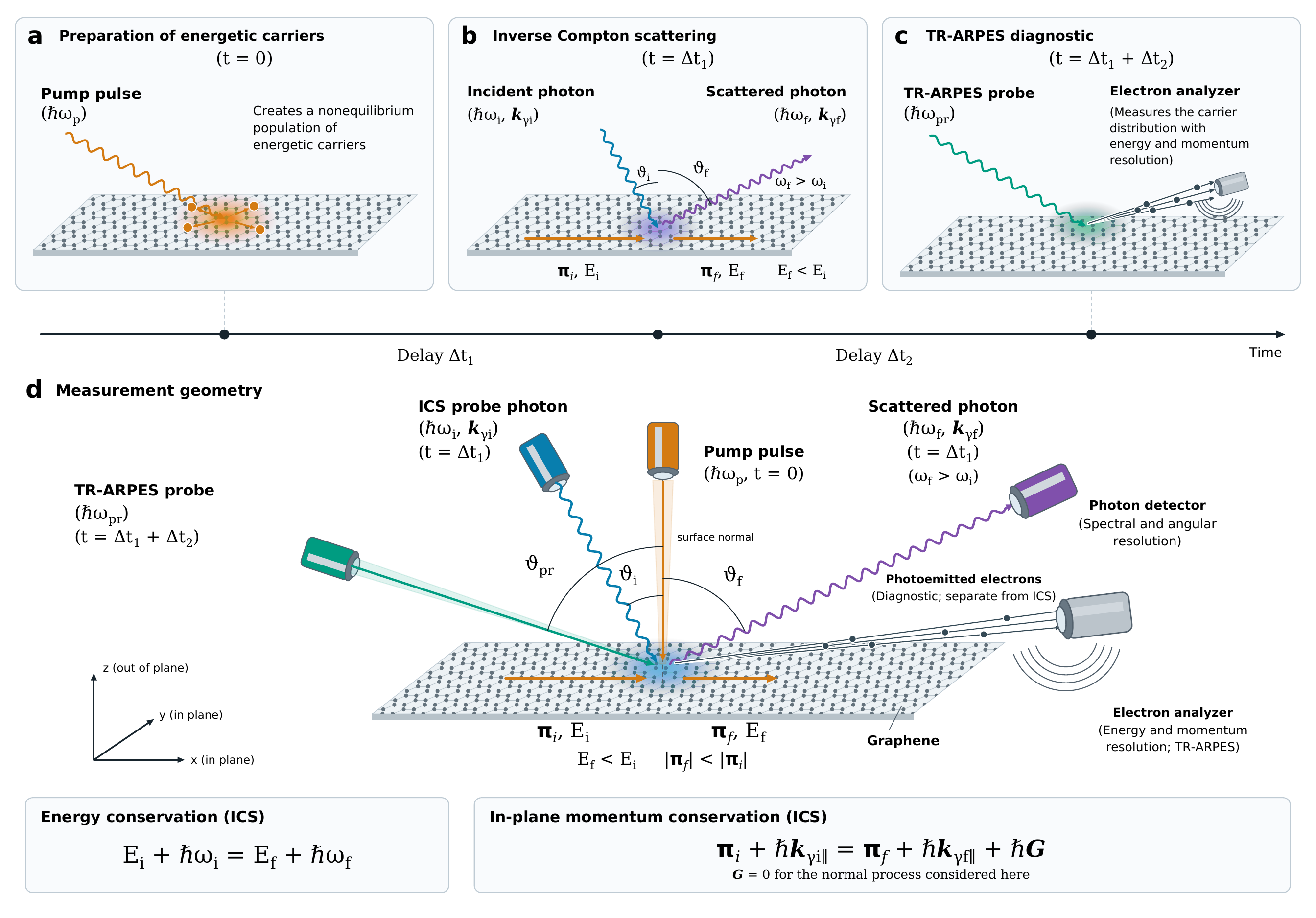}%
}{%
    \fbox{\parbox[c][5.5cm][c]{0.95\textwidth}{\centering
    Figure file: \texttt{ics\_trarpes\_schematic\_pi\_updated.pdf}\\
    Place the original figure PDF beside this source file.}}%
}
    \caption{Proposed time-resolved geometry for inverse Compton scattering (ICS) in graphene and the independent diagnosis of the transient carrier distribution using time-resolved angle-resolved photoemission spectroscopy (TR-ARPES). (a) Preparation of energetic carriers: an ultrafast pump pulse of energy $\hbar\omega_p$ excites graphene at $t=0$, creating a nonequilibrium carrier population. (b) Inverse Compton scattering: at $t=\Delta t_1$, an incident photon of energy $\hbar\omega_i$ and wave vector $\mathbf{k}_{\gamma i}$ interacts with the carriers, producing a scattered photon of energy
$\hbar\omega_f$ and wave vector $\mathbf{k}_{\gamma f}$, with
$\omega_f>\omega_i$ and $E_f<E_i$. (c) TR-ARPES diagnostic: at $t=\Delta t_1+\Delta t_2$, a separate
probe pulse of energy $\hbar\omega_{\mathrm{pr}}$ photoemits electrons whose energy and momentum are analysed to determine the transient carrier distribution. These photoelectrons are not products of the ICS process. (d) Measurement geometry: schematic arrangement of the pump pulse, incident ICS photon, scattered-photon detector, TR-ARPES probe, and electron analyser, together with the relevant incidence and emission angles and the in-plane coordinate directions. The lower panels also show the energy-conservation relation for ICS and the in-plane momentum-conservation relation, including the reciprocal-lattice vector $\mathbf{G}$;  $\mathbf{G}=\mathbf{0}$ for the normal process considered here.}
\label{fig:graphene_ics_trarpes_geometry}
\end{figure}

The proposed measurement sequence is shown in
Fig.~\ref{fig:graphene_ics_trarpes_geometry}. An ultrafast pump pulse
first prepares the nonequilibrium carrier population. After a delay
$\Delta t_1$, the incident photon interacts with the carriers, and
the scattered radiation is recorded. A separate TR-ARPES probe pulse
then examines the carrier distribution after a further delay
$\Delta t_2$. Thus, $\Delta t_1$ is the pump-to-ICS delay, while
$\Delta t_2$ is the ICS-to-TR-ARPES delay. Both delays must be chosen
with the carrier-scattering and relaxation times in mind.

The proposed ICS process should exhibit photon up-conversion
accompanied by a loss of carrier energy:
\(\omega_f>\omega_i,\quad E_f<E_i.
\label{eq:ics_energy_transfer_signature}
\)
Here $E_i$ and $E_f$ are the initial and final carrier energies.
Energy conservation requires
\(E_i+\hbar\omega_i=E_f+\hbar\omega_f.
\)
This gives
\( E_i-E_f =\hbar(\omega_f-\omega_i)>0.\)
The increase in photon energy must therefore be balanced by a
corresponding decrease in carrier energy \cite{Serafini2024}.

Energy conservation alone, however, is not sufficient to identify
the process. The in-plane momentum transfer must also be consistent
with the scattering event. Let $\boldsymbol{\pi}_i$ and
$\boldsymbol{\pi}_f$ denote the initial and final electronic crystal
momenta measured relative to the relevant Dirac point. The
corresponding conservation relation is
\(\boldsymbol{\pi}_i+\hbar\mathbf{k}_{\gamma i,\parallel}
=\boldsymbol{\pi}_f+\hbar\mathbf{k}_{\gamma f,\parallel}
+\hbar\mathbf{G}.
\)
Here $\mathbf{k}_{\gamma i,\parallel}$ and $\mathbf{k}_{\gamma f,\parallel}$ are
the in-plane photon wave vectors, and $\mathbf{G}$ is a reciprocal-
lattice vector. In the single-valley continuum approximation,
when Umklapp scattering is neglected, $\mathbf{G}=0$. The energy
and momentum relations must be satisfied simultaneously using
the appropriate graphene dispersion.

The importance of the interaction geometry is illustrated by the
shallow-angle inverse Compton scattering experiment of Schaap
\textit{et al.}, in which the incidence geometry affected the
spectrum, angular distribution, and polarization dependence of the
emitted radiation \cite{Schaap2026}. This work provides motivation
for exploring angle-dependent photon scattering in graphene.
Nevertheless, the kinematics of a free relativistic electron cannot
be applied directly to graphene, where the carrier dispersion and
in-plane momentum conservation must be treated explicitly.

Another challenge is to distinguish the proposed ICS process from
other mechanisms that can produce higher-frequency radiation,
including harmonic generation, frequency mixing, and photon-drag-
related processes \cite{Glazov2014,Ornigotti2023}. Useful control
measurements would include experiments without the incident probe
beam, without energetic-carrier preparation, and with an appropriate
reference sample. The dependence of the scattered spectrum on
carrier energy, incidence angle, and polarization would provide
further tests of the proposed mechanism.

TR-ARPES could strengthen this analysis by revealing how the
nonequilibrium carrier distribution changes with time. Comparing
these measurements with the photon spectra would help determine
whether the observed electronic evolution is consistent with the
energy and momentum transferred to the scattered radiation.
However, TR-ARPES measures the carrier population collectively and
cannot, by itself, establish that a particular carrier participated
in an ICS event.

The proposed experiment thus combines established methods for
preparing and characterizing nonequilibrium carriers in graphene
\cite{Malic2017,Gierz2017} with a dedicated photon-scattering
measurement. Evidence for ICS would require more than observing
photon up-conversion: the measured radiation should follow the
predictions of graphene-specific scattering kinematics and respond
systematically to changes in the carrier distribution and interaction
geometry. Where feasible, independent measurements of the carrier
distribution would provide an additional check. Together, these
measurements could help test the proposed photon--Dirac-carrier
scattering process and distinguish it from competing nonlinear
optical mechanisms.
\section{Conclusions}

We have formulated the kinematics of Compton and inverse Compton scattering
from Dirac carriers in graphene. The central distinction from the vacuum
problem is the coexistence of the graphene velocity $\vF$ and the
electromagnetic velocity $c$. Because $\vF\ll c$, the vacuum Compton
four-momentum construction cannot be transferred to graphene by simply
replacing $c$ with $\vF$.

For a gapless Dirac carrier initially at the Dirac point, simultaneous energy
and in-plane momentum conservation yields Eq.~\eqref{eq:dirac_point_relation},
which describes photon energy loss rather than inverse-Compton up-conversion.
For gapped graphene, the stationary-carrier problem remains nontrivial at
finite scattering angle, but the opening of a gap does not restore the vacuum
Lorentz invariant and does not by itself produce an inverse-Compton channel.

The principal result is the exact quadratic relation
Eq.~\eqref{eq:graphene_ics_quadratic} for an energetic gapless Dirac carrier.
Its physical solution, Eq.~\eqref{eq:graphene_ics_solution}, provides the
appropriate kinematic relation for the proposed graphene ICS process. Genuine
inverse Compton scattering requires $\omega_f>\omega_i$ together with
$E_f<E_i$ and the corresponding in-plane momentum balance.

Finally, we have proposed a pump--probe measurement in which the scattered
photon is the primary ICS observable and TR-ARPES provides an independent
diagnostic of the transient carrier distribution. Appropriate control
measurements are essential because graphene supports several alternative
frequency-conversion mechanisms. The present work is therefore a kinematic
framework and experimental proposal, not an established observation.
\begin{acknowledgments}
An artificial-intelligence image-generation tool was used to assist in
preparing the schematic illustration in Fig.~2. The figure was directed,
edited, and scientifically verified by the author, who is fully responsible
for its accuracy and content.

\end{acknowledgments}

\end{document}